\input harvmac
\noblackbox
%
%

\def\NP{{\it Nucl. Phys.\ }}

\def\CMP{{\it Comm. Math. Phys.\ }}

\font\tiau=cmcsc10
\baselineskip 12pt
\Title{\vbox{\baselineskip12pt \hbox{}
\hbox{} }}
{\vbox{\hbox{\centerline{\bf GENERALIZED MATHAI-QUILLEN TOPOLOGICAL 
SIGMA MODELS
}}}}
\centerline{\tiau Pablo M. Llatas\foot{llatas@denali.physics.ucsb.edu}}

\vskip.1in
\centerline{\it Department of Physics}
\centerline{\it University of California}
\centerline{\it Santa Barbara, CA 93106-9530}
\vskip .9cm
\noindent{A simple field theoretical approach to Mathai-Quillen topological 
field 
theories of maps $X: M_I\rightarrow M_T$ from an internal space to a target
space is presented.
 As an example of applications of our formalism we compute
by applying our formulas the action and $Q$-variations of the fields of two
well known topological systems: Topological Quantum
Mechanics and type-A topological Sigma Model.  }

\Date{July, 1996}

\lref\wittena{E. Witten,  \CMP {\bf 117} (1988)353.}

\lref\wittenb{E. Witten,  \CMP {\bf 118} (1988) 411.}

\lref\wittenc{E. Witten, ``Mirror Manifolds and Topological Field 
Theory'', in Essays on Mirror Manifolds, International Press (1992)
(hep-th/9112056).}

\lref\blaua{M. Blau, ``The Mathai-Quillen Formalism and Topological 
Field Theory'', hep-th/9203026.}

\lref\labastida{J.M.F. Labastida, \CMP {\bf 123} (1989) 641.}

\lref\yo{J.M.F. Labastida and P.M. Llatas, \NP {\bf B271} (1991) 101.}

\lref\moore{S. Cordes, G. Moore and S. Ramgoolam, hep-th/9411210.}

\lref\blaub{D. Birmingan, M. Blau, M. Rakowski and G. Thompson, 
Phys. Rep. {\bf 209} (1991) 129.}

\lref\mq{V. Mathai and D. Quillen, Topology {\bf 25} (1986)85.}

\lref\nash{C. Nash  ``Differential Topology and Quantum Field Theory'', 
Academic Press 1991.}

\baselineskip 12pt

\newsec{Introduction}

Topological Field Theories (TFT's) have been extensively studied
during the past recent years (see \refs{\blaub} and references therein). 
The topological sigma models first introduced in \refs{\wittenb} were
motivated by the idea of the possibility of understanding gravity as a 
broken phase 
of a topological theory. In particular, the string theory would be
a broken phase of a topological sigma model. This topological sigma
model was obtained in \refs{\wittenb} by the so-called 
topological twisting of a $N=2$
supersymmetric sigma model, where the internal manifold was bidimensional
($d=2$).
This formalism requires the target 
manifold to be a Kahler manifold (in order to have $N=2$ supersymmetry).
However, Witten showed that the kahler condition could be relaxed 
to Hermitean target manifolds. In the present work we introduce a technique
to obtain topological sigma models different from that of twisting 
a supersymmetric theory. This allow us to write topological actions
in which the internal manifold can have any dimension $d=m$ and 
the target manifold can be any smooth manifold (even a real manifold), 
what constitutes a considerable generalization. The condition that we are going
to impose from the begining is the localization of the correlators of
the theory on a certain moduli space $\cal M$ of instantons. Is for 
this reason that we have called the theories so obtained    
Mathai-Quillen Topological Sigma Models (MQTSM). As we also show, 
the topological 
sigma model of type A (\refs{\wittenc ,\yo}) and the topological 
quantum mechanics ({\refs\labastida}) are particular cases of MQTSM.
Mathai-Quillen \refs{\mq} topological field theories has been previously
studied in \refs{\blaua} and \refs{\moore}. Our approach is more
field-theoretical and less geometrical that the ones presented there.

In section two we introduce our definition of Mathai-Quillen Topological
Sigma Models. In section three we will introduce a fermionic transformation
${\nabla}_Q$ (closely related to the transformation ${\delta}_Q$ associated
to the topological charge $Q$ present on any TFT) which simplifies notably
the study of the geometrical properties of the theory in the target space.
Important properties of ${\nabla}_Q$ are discussed there. In section four
we construct the most general ``basic'' action of the MQTSM type and give
the ${\delta}_Q$-transformations of the fields. 
We also comment there the topological character of the theory 
and discuss the observables and correlators of MQ field theory (leading to
intersection numbers on a moduli space of instantons). In section five we
apply our formulas to two known examples, the Topological Quantum Mechanics
(\refs{\labastida}) and the type-A Topological Sigma Models (\refs{\wittenb
,\yo}). Finally, in section six we present our conclusions.

\newsec{The Generalized Mathai-Quillen Topological Sigma Models.}

In this section we will introduce the elements defining a Mathai-Quillen
Topological Sigma Model (MQTSM).
\vskip .5cm 

$\bullet$ First we introduce two smooth $C^{\infty}$ differentiable
manifolds (one m-dimensional internal manifold $M_I$ and one n-dimensional
target manifold $M_T$), together with a continuous map $X$:
\eqn\map{
X :M_I\longrightarrow M_T.  }   
Given an atlas $\cup_a ({\cal U}_a
,{\sigma}^{\mu}_{(a)})$ on $M_I$ and an atlas $\cup_A ({\cal U}_A
,X^i_{(A)})$ on $M_T$, the functions $X^i_A (\sigma^{\mu}_{a})$
are $C^{\infty}$-functions in the $\sigma^{\mu}_{(a)}$
coordinates (here ${\cal U}_a$ (${\cal U}_A$) are the open subsets of a
covering on $M_I$ ($M_T$) and ${\sigma}^{\mu}_{(a)}$, $\mu :1,....,m$
($X^i_{(A)}$, $i:1,....n$) are a system of local coordinates on ${\cal U}_a$
(${\cal U}_A$)). From now on we will drop the covering indices $a$ and $A$
to simplify notation. The functions $X^i(\sigma^\mu)$ ($X^i: U^m\rightarrow
U^n$, being $U^m$ and $U^n$ open subsets of $R^m$ and $R^n$ respectively) 
give the coordinates of $M_I$ as immersed on $M_T$.  Further, we
provide $M_I$ and $M_T$ with Euclidean metrics $g_{\mu\nu}$ and $G_{ij}$
respectively. We will denote by $G_{ij}(X)$ the restriction of the metric
$G_{ij}$ to the submanifold $X^i (M_I)\subset M_T$.  Under a change of
coordinates on $M_I$ (${\sigma '}^{\mu}= {\sigma '}^{\mu} ({\sigma})$) the
functions $X^i (\sigma)$ behave as scalars:
\eqn\cc{
{X'}^i ({\sigma '})= X^i ({\sigma}) }
Under a change of coordinates in
the target the functions $X^i (\sigma)$ change as usual:
\eqn\cct{
{X'}^i (\sigma)={X'}^i (X(\sigma))\sim X^i (\sigma) +\xi^i(X(\sigma)) }
(where in the last step we have made the coordinate transformation
infinitesimal).
\vskip .5cm

$\bullet$ Our second important ingredient for MQTSM formalism (like in any
other TFT model) is the existence of a fermionic symmetry transformation
$\delta_Q$ generated by a fermionic operator $Q$.  By symmetry we mean that
the action ($S[\Phi]$), observables (${\cal O}(\Phi)$) and measure (${\cal
D}\Phi$) of the theory are invariant under $Q$:
\eqn\ini{
\delta_Q (S[\Phi]) =\delta_Q ({\cal D}\Phi ) =\delta_Q ({\cal O}(\Phi)) =0
} (here $\Phi$ denotes the field space of the theory). This fermionic
operator is taken to be a scalar on $M_I$ (this fixes the properties of 
$Q$ under target changes of coordinates that 
will be analyzed in the next section).
Moreover, we take this operator to be nilpotent on the field space
$\Phi$:
\eqn\ni{
\delta_Q^2 \Phi =0 
} Finally, we demand the internal metric $g_{\mu\nu}(\sigma)$ to be
invariant under $Q$:
\eqn\im{
\delta_Q g_{\mu\nu}(\sigma)=0
} (in particular, this means that our model is not coupled to topological
gravity).
\vskip .5cm
$\bullet$ Our third request is that the action $S[\Phi]$ is taken to be
$Q$-exact:
\eqn\qac{
S_t[\Phi]=t \delta_Q (A[\Phi])\equiv tS[\Phi] } 
for some scalar functional $A[\Phi]$ ($t$
is a c-number parameter). The previous equation has an important
consequence (\refs{\wittenb}). The correlators of
the theory are independent of the parameter $t$:
\eqn\co{
<{\cal O}_1\ldots {\cal O}_p>_t =\int{[{\cal D}\Phi]\,\, {\cal O}_1
(\Phi)\ldots {\cal O}_2 (\Phi) e^{i S_t [\Phi]}} }
\eqn\ll{
-i{d\over dt}<{\cal O}_1 \ldots {\cal O}_p>_t= <{\cal O}_1\ldots {\cal O}_p
S[\Phi]>_t =<\delta_Q ({\cal O}_1\ldots {\cal O}_p A[\Phi])>_t=0 } (in the
last step we have used that $Q$ is an exact symmetry of the quantum
theory). This observation is very important because allows us to compute
correlators in the most convenient value of $t$ (typically one makes the
limit $t\rightarrow\infty$ where the computation of the path integral
reduces to the semiclassical limit). We will make use of this property
when discussing the topological character of these theories in section 
four.

\vskip .5cm
$\bullet$ The fourth and final property of our definition of MQTSM is the
one that gives the name ``Mathai-Quillen'' to the MQTSM's. We will demand
the correlators of the theory to be localized on a certain moduli space
$\cal M$ given by (\refs{\moore}):
\eqn\moduli{
{\cal M}= \{\phi\subset\Phi \mid s=D\phi =0\}/G } where $\Phi$ is again the
space of fields, $D$ is some chosen differential operator, $s=D\phi$ is a
section of a vector bundle over $M_T$ and $G$ is some group of 
symmetries present in the theory. In other words, we want the functional
integrals defining the correlators to be localized on the subspace $\phi$
of the field space $\Phi$ satisfying $D\phi=0$ (modulo symmetry 
 transformations).
This localization is obtained by constructing an action of the form:
\eqn\loc{
S_t[\Phi]=t \delta_Q (A[\Phi])=t (\| D\phi\|^2+\ldots ) } then, it is
immediate to see from \co\ that, in the large $t$ limit, 
we get localization on ${\cal M}$:
\eqn\loca{
\lim_{t\to\infty} <{\cal O}_1\ldots {\cal O}_p>_t =
<{\cal O}_1\ldots {\cal O}_p>_{\cal M} } We will refer sometimes to $\phi$
as the ``instantons'' and $\cal M$ as the moduli space of instantons.

\newsec{The operator $\nabla_Q$.}

The only field of the field space $\Phi$ that we have specified so far are
the fields $X^i (\sigma)$ ($i:1,..,n$). Using \ni\ we can deduce the next
relations (from now on we will drop the internal coordinates $\sigma^\mu$
and only write them if we consider it clarifying):
\eqn\re{
\delta_Q X^i=\chi^i \,\, ;\qquad\qquad \delta_Q \chi^i =0.
} Being $Q$ a fermionic operator and the $X^i$ independent fields we have
that the $\chi^i$ are $n$ independent fermionic fields. The rest of this
section will be devoted to the analysis of the simple relations \re . The
first observation is that, being both $X^i$ (see \cc ) and $Q$ scalar
objects from the point of view of $M_I$, the fields $\chi^i$ are also
scalar fields with respect to change of coordinates on the internal
manifold. Now, let us study the behaviour of $\chi^i$ under change of
coordinates in the target manifold $M_T$. From \cct\ and
\re\ we get that under an infinitesimal change of coordinates on $M_T$:
\eqn\vc{
{\chi '}^i =\delta_Q {X'}^i \sim \chi^i +\partial_j \xi^i (X)\chi^j } i.e.,
we conclude that the fields $\chi^i$ ($i:1,...,n$) are the components of a
vector in the target. Note that this is not a new condition, but a
consequence of \cct\ and \re . An implication of this observation is that
the $Q$ operator is not a (scalar) covariant operator from the point of
view of the target space $M_T$. This is so because $Q$ acting on target
coordinate fields ($X^i$) produces vector-target fields ($\chi^i$). This
introduces some problem when analyzing the geometrical aspects of the
theory on the target. Let us analyze the problem more carefully. Let us
take an arbitrary 
 target-vector $V^i (X)$. Under a change of coordinates like \cct\ we
have:
\eqn\vc{
{V}^i (X)\rightarrow {V'}^i (X')\sim V^i (X)+ \partial_j \xi^i (X) V^j (X)
} therefore, applying \re\ we get:
\eqn\rev{
\delta_Q {V'}(X')\sim \delta_Q V^i (X)+\partial_j \xi^i (X)\delta_Q 
V^j (X) +\partial_k\partial_j \xi^i (X)\chi^j V^k } 
We observe then that
$\delta_Q V^i (X)$ is not a target vector due to the last term in \rev\ .
Now, using the target metric $G_{ij}$ we construct the affine
connection $\Gamma^{i}_{jk}$ (we will consider torsionless connections).
One easily verifies that under an infinitesimal change of coordinates the
object $O^i (X)\equiv \Gamma^i_{jk}(X)V^j_1 (X) V^k_2 (X)$ (here $V^i_1
(X)$ and $V^i_2 (X)$ are two arbitrary target vectors) transforms as:
\eqn\ch{{O'}^i (X')\sim O^i (X) +\partial_j \xi^i (X)O^j (X) -
\partial_k\partial_j \xi^i (X)V^k_1 (X)V^j_2 (X).
} Comparing now with \rev\ we are led to naturally introduce the
transformation:
\eqn\nablaa{
\nabla_Q V^i (X)=\delta_Q V^i (X)+\Gamma^i_{jk}(X)\chi^j V^k (X)=
\chi^j D_j V^i (X)
} which maps target-vectors into target-vectors. $\nabla_Q$, contrary to
$\delta_Q$, is then a covariant scalar in the target.  This analysis was
done for vector fields depending only on $X^i$ (like $V^i (X)$). However,
we generalize our definition to vector fields depending on any field
$\phi\subset\Phi$. Then we define the $\nabla_Q$ transformation by (the
analysis for the covariant case can be done in the same way):
\eqn\nablag{
\nabla_Q V^i (\phi)\equiv\delta_Q V^i (\phi)+\Gamma^i_{jk}(X)\chi^j V^k (\phi)
}
\eqn\nablagb{
\nabla_Q V_i (\phi)\equiv\delta_Q V_i (\phi)-\Gamma^k_{ij}(X)\chi^j V_k (\phi)
}
We note here that, for a general field $\phi$ 
(contrary to the last relation of \nablaa\ for the special case $\phi =X$)
$\nabla_Q V^{i} (\phi)=\delta_Q V^i (\phi )+\Gamma^i_{jk}\chi^j V^k (\phi )
\ne \chi^k D_j V^i (\phi)$. Also, we remark that, due to the fact that $\Phi$
contains fermionic fields, the position of the $\chi^i$ fields in our
previous definitions are important.  The generalization of our definitions
to tensors with any number of covariant and contravariant indices is
trivial.  The next properties of the $\nabla_Q$ transformation can be
straightforwardly checked:
\eqn\bull{
\eqalign{
&\bullet \nabla_Q (A(\Phi)B(\Phi))=(\nabla_Q A(\Phi))B(\Phi)+
(-)^{\epsilon_A} A(\Phi)(\nabla_Q B(\Phi)).\cr 
&\bullet \nabla_Q G_{ij} (X)=0.\cr
&\bullet \nabla_Q \chi^i=\delta_Q \chi^i =0.\cr 
&\bullet \nabla_Q (A^i
(\Phi)B_i (\Phi))=\delta_Q (A^i (\Phi)B_i (\Phi)).\cr 
&\bullet \nabla^2_Q
A^i (\Phi)={1\over 2}R_{jkl}^{i}(X) \chi^j\chi^k A^l (\Phi).\cr} } 
(the last relation holds if $\nabla_Q A^i (\phi)\neq 0$). Here,
$\epsilon_A$ is $0$ ($1$) if $A(\Phi)$ is a bosonic (fermionic) operator.
$R_{jkl}^{i}(X)$ is the curvature tensor on the target space.  A very
important observation should be made from these relations.  Looking at the
fourth relation in \bull\ we have that the $\nabla_Q$-transformation on any
scalar is $Q$-exact ($\nabla_Q =\delta_Q$ when acting on target-scalars).
In particular, the equations \ini\ and \qac\ implies that:
\eqn\ma{
\nabla_Q (S[\Phi])=\nabla_Q ({\cal D}\Phi )=\nabla_Q ({\cal O}[\Phi])=0
} and
\eqn\mb{
S_t [\Phi]=t\delta_Q (A[\Phi])=t\nabla_Q (A[\Phi]).  } 
This means that
$\nabla_Q$ defines also a symmetry of the quantum theory. Finally, 
let us notice that contrary to $\delta_Q$, $\nabla_Q$ is not 
a nilpotent operator.

\newsec{The Action and $Q$-transformations for MQTSM's.}

In this section, with minimal information, we will construct a rather
general action localizing the correlators in some chosen moduli space
$\cal M$.  By minimal information we mean that we will not specify the
field content of the theory ($\Phi$). To construct concrete examples later,
we will have to be more specific in this aspect.  Also we write the
$\delta_Q$-transformations and $\nabla_Q$-transformations that with this
minimal data can be analyzed. A discussion of the topological character 
and observables of the theory is also presented.

\subsec{The Action for MQTSM's.}

We already have lot of information for constructing an action $S[\Phi]$ for
MQTSM's.  First, we know that the action is $\nabla_Q$-exact \mb\ , and
second, we want $S[\Phi]$ to depend on a given bosonic section $s[\phi]$
($\phi\subset\Phi$) of a vector bundle over $M_T$. The most simple term to
start with is:
\eqn\acta{
S^{1}_{t} [\Phi]=t \nabla_Q 
(\int{d^m \sigma \sqrt{g(\sigma)} \rho^{*}_{i} [\Phi] s^{i}_{*}
[\Phi]})=
\int{\sqrt{d^m \sigma g(\sigma)}  
((\nabla_Q \rho_{i}^{*} [\Phi])s^{i}_{*} [\Phi]- 
\rho_{i}^{*} [\Phi]
(\nabla_Q s^{i}_{*} [\Phi]))} } Here, $\rho_{i}^{*} [\Phi]$ is any
fermionic function of the field space $\Phi$ (the action has to be
bosonic).  $s^{i}_{*} [\Phi]$ is the bosonic section of the vector bundle
and ``$*$'' denotes all internal indices (like indices associated to the
internal manifold $M_I$ or gauge indices) which are conveniently contracted
to make $S_t [\Phi]$ a scalar functional. To avoid complicate notation we will
also drop out the ``$*$'' from some of the expressions and only restore them
whenever we will consider it clarifying. The action that we have obtained
in \acta\ is not gaussian in the section as we wish (see \loc ). 
But looking at it it is
easy to guess what we have to do to get such a gaussian term. Just introduce
a ``metric'' $A^{**}_{ij}$ such that
$\rho_{i}^{*}[\Phi]=A^{**}_{ij}[\Phi]\rho^{i}_{*}[\Phi]$, and add a term
$\nabla_Q (\int{ \sqrt{g(\sigma)} d^m\sigma 
(\rho_{i}^{*}\nabla_Q \rho^{i}_{*})})$ to the
action. One finds:
\eqn\pablo{
S_t [\Phi]=t \nabla_Q \Bigl( \int{d^m \sigma\sqrt{g(\sigma)} 
(\rho^{*}_{i} [\Phi] s^{i}_{*}
[\Phi]}+\rho_{i}^{*}[\Phi]\nabla_Q \rho^{i}_{*}[\Phi])\Bigr).
}
Using the properties in \bull : 
\eqn\actb{
\eqalign{
S_t[\Phi]=&t\int d^m \sigma \sqrt{g(\sigma)} \Bigl\{ A_{ij}[\Phi]
\bigl( (\nabla_Q \rho^i [\Phi]) s^j
[\Phi]- \rho^i [\Phi] \nabla_Q (s^j [\Phi]) + (\nabla_Q\rho^i [\Phi]
)(\nabla_Q\rho^j [\Phi] )\cr 
&\qquad\qquad\qquad - \rho^i [\Phi] \nabla_{Q}^{2} \rho^j [\Phi]\bigr)
+(\nabla_Q A_{ij}[\Phi])(\rho^i [\Phi]s^j [\Phi]+\rho^i [\Phi]\nabla_Q 
\rho^j [\Phi]) \Bigr\} =\cr 
=& t\int d^m \sigma\sqrt{g(\sigma)} \Bigl\{ A_{ij} [\Phi]
\bigr( (\nabla_Q \rho^i [\Phi] + {1\over
2}s^i [\Phi])(\nabla_Q \rho^j [\Phi] +{1\over 2}s^j [\Phi])- {1\over 4} s^i
[\Phi] s^j [\Phi]\cr 
&\qquad\qquad\qquad - \rho^i [\Phi] \nabla_Q s^j [\Phi] 
-{1\over 2}R^{j}_{klm} (X)\chi^k \chi^l\rho^i [\Phi]\rho^m [\Phi]\bigr)\cr
&\qquad\qquad\qquad 
+(\nabla_Q A_{ij}[\Phi])(\rho^i [\Phi]s^j [\Phi]+\rho^i [\Phi]\nabla_Q 
\rho^j [\Phi])\Bigr\}.\cr}
}
Defining the ``auxiliary'' fields $H^{i}_{*}$ by:
\eqn\h{
H^{i}_{*} [\Phi]\equiv \nabla_Q\rho^i_{*} [\Phi]+{1\over 2}s^i_{*} [\Phi].
}
we finally get the action:
\eqn\tth{
\eqalign{
S_t[\Phi]=&t\int d^m \sigma\sqrt{g(\sigma)} \Bigl\{ A_{ij} 
[\Phi]\bigl( (H^i [\Phi]H^j [\Phi]-
{1\over 4}s^i [\Phi] s^j [\Phi] -\rho^i [\Phi] \nabla_Q s^j [\Phi] \cr
&\qquad\qquad\qquad 
-{1\over 2}R^{j}_{klm} (X)\chi^k \chi^l \rho^i [\Phi]\rho^m [\Phi]\bigr)\cr
&\qquad\qquad\qquad 
+(\nabla_Q A_{ij}[\Phi])(\rho^i [\Phi]s^j [\Phi]+\rho^i [\Phi]\nabla_Q 
\rho^j [\Phi])\Bigr\} .\cr }
} 
In \tth\ we have obtained a desired action like the one in \loc\ if the
tensor $A_{ij}^{**} [\Phi]$ defines a proper norm (i.e., defines a
positive-definite quadratic form):
\eqn\norm{
\|D\Phi \|^2=A^{**}_{ij}[\Phi]s^i_{*} [\Phi]s^j_{*} [\Phi].
} Using the $t\rightarrow\infty$ argument we see that the path integration
of a field theory with the previous action is localized on configurations
with $s^{i}_{*} [\Phi]=0$ as we want.  
Our analysis is rather general (we have not
specify neither the form of the section $s^{i}_{*} [\Phi]$ nor the form of
the function $\rho_{i}^{*} [\Phi]$), but this is the ``minimal'' form of
the action for MQTSM's.  ``Non-minimal'' actions can be obtained by adding
$\nabla_Q$-exact (i.e., $Q$-exact) terms to this ``minimal'' action.

\subsec{Basic $Q$-Transformations}

To write the precise $Q$-transformations of the fields in the theory
requires to specify $\Phi$ and the concrete form of $s^i_{*} [\Phi]$ and
$\rho^i_{*} [\Phi]$.  Nevertheless we can already get valuable general
information over the structure of such transformations and write
expressions for them.  Here we collect such expressions:
\eqn\nqt{
\eqalign{
&\bullet \delta_Q X^i =\chi^i\cr
&\bullet \nabla_Q \chi^i =0.\cr &\bullet \nabla_Q \rho^{i}_{*} [\Phi]
=H^{i}_{*} [\Phi]-{1\over 2} s^{i}_{*}[\Phi]\cr 
&\bullet \nabla_Q H^{i}_{*}
[\Phi] ={1\over 2}\nabla_Q s^{i}_{*} [\Phi] +{1\over 2}R^{i}_{jkl}(X)
\chi^j\chi^k\rho^{l}_{*} [\Phi]\cr }
}
The first two relations were already known in \re\ and \bull\ and the third in
\h . Finally, the last one is obtained trivially by applying 
$\nabla_Q$ to the third and using the last property of \bull . 
From these relations \nqt\ we can derive straightforwardly the 
$\delta_Q$-transformations, just by using the definition \nablag .
One gets:
\eqn\qt{
\eqalign{
&\bullet \delta_Q X^i =\chi^i.\cr 
&\bullet \delta_Q \chi^i =0.\cr 
&\bullet
\delta_Q \rho^{i}_{*} [\Phi] =H^{i}_{*} [\Phi]-{1\over 2} s^{i}_{*}[\Phi]
-\Gamma^{i}_{jk}(X)\chi^j\rho^k_{*} [\Phi]\cr 
&\bullet \delta_Q H^{i}_{*}
[\Phi] ={1\over 2}\delta_Q s^{i}_{*} [\Phi] + {1\over
2}\Gamma^{i}_{jk}(X)\chi^j s^k_{*} [\Phi] -\Gamma^{i}_{jk}(X)\chi^{j}
H^{k}_{*} [\Phi]\cr 
&\qquad\qquad\qquad +{1\over 2}R^{i}_{jkl}(X)
\chi^j\chi^k\rho^{l}_{*} [\Phi]\cr }
}
One checks that the $\delta_Q$-transformations 
above are automatically nilpotent (as demanded by \ni ).

\subsec{Topological Character}

Let us study the topological character of the 
theory so far presented. We do not know a rigorous proof  
to justify the topological character of the theory neither on 
the internal manifold $M_I$ nor on the target $M_T$
holding at any value of the parameter $t$. We have an 
argument that applies in the large $t\rightarrow\infty$ limit and then 
we use \ll\ to generalize it for any value of $t$.
Let us first study the behaviour of the 
theory under deformations of the metric in the internal manifold 
$g_{\mu\nu}$. 
The $\delta_Q$-exactness of the action \tth\   
is not enough to guarantee the invariance of the theory under 
deformations of the metric. This is so due to the possible metric dependence 
of the transformations \qt\ (this dependence could appear in 
$s^i_{*} [\Phi]$, $\rho^i_{*} [\Phi]$ and $H^i_{*} [\Phi]$). 
Actually, only in the case in which the 
$\delta_Q$-transformations are independent of the metric we have 
that deformations of the internal metric $\delta g_{\mu\nu}$ and 
$\delta_Q$-transformations commute and, therefore, that the energy-momentum 
tensor $T_{\mu\nu}$ is $Q$-exact. General arguments shows that $Q$-exact
energy momnetums lead to topological theories (\refs{\wittena})
However, if we assume that all the metric dependence in the 
$\delta_Q$-transformations \qt\ are in the section $s^i [\Phi]$ 
and the auxiliary field $H^i_{*} [\Phi]$ (as in the cases that we will 
consider), 
then we see that the transformations  
\qt\ are independent of the metric $g_{\mu\nu}$ if we restrict ourselves to 
the moduli space $s^i [\Phi]=H^i [\Phi]=0$
($i.e.$, $\cal M$), where the path integral is localized in the 
limit $t\rightarrow\infty$ (thanks to the $\delta_Q$-exactness of the 
action we can take that limit as correct). We conclude then that, in this
case,
deformations of $g_{\mu\nu}$ and $\delta_Q$-transformations commute 
and then   
the theory described by the action \tth\ is topological with respect 
to the internal manifold $M_I$. In fact, in the examples to be consider 
later we will take $\rho^i_{*} [\Phi]$ and $H^i_{*} [\Phi]$ to be 
independent elemental fields ($\rho^i_{*}$ and $H^i_{*}$ respectively), 
and we have then a topological field theory with respect to $M_I$.

The analysis of the topological character on the target space $M_T$ is 
similar. If we observe the $\nabla_Q$-transformations in \nqt\ 
we deduce that in the case in which all the dependence on $G_{ij}$ in \nqt\ 
is in $s^i [\Phi]$ and in $H^i [\Phi]$ then, in the $t\rightarrow\infty$
limit ($i.e.$, in the moduli space $\cal M$ defined by $s^i_{*}[\Phi]=
H^i_{*}[\Phi]=0$), all the dependence of the $\nabla_Q$-transformations 
in the target metric is in the 
curvature term in the transformation of the field $H^i_{*} [\Phi]$. 
But is easy to check that, on $\cal M$
(here, $\hat{\delta}$ means deformations with respect to the target metric 
$G_{ij}$):
\eqn\chapuza{
\eqalign{
\lim_{t\to\infty} \hat{\delta} \Bigr( 
\nabla_Q H^i_{*}[\Phi]\Bigl) &=\lim_{t\to\infty} 
{1\over 2}\hat{\delta}( R^i_{jkl})\chi^j\chi^k\rho^l_{*} [\Phi]=
\lim_{t\to\infty} {1\over 2}\hat{\delta}\Big\{\delta_Q
\Bigr( \Gamma^i_{kl}\chi^k\rho^l_{*} [\Phi]\Big)\Big\} \cr
&=
\lim_{t\to\infty} {1\over 2}\nabla_Q\Big\{\hat{\delta}
\Bigr( \Gamma^i_{kl}\chi^k\rho^l_{*} [\Phi]\Big)\Big\}.}
}
We have used that $\hat{\delta}$ commute with $\delta_Q$ when acting on $X^i$ 
and $\chi^i$. Also note that, although $\Gamma^i_{jk}$ is not a tensor, 
$\hat{\delta}\Gamma^i_{jk}$ is a tensor and, therefore, the $\nabla_Q$ 
action in the last term in \chapuza\ is wel defined. 
In \chapuza\ we see that deformations $\hat{\delta}$ of the target metric  
varies the $\nabla_Q$-transformations by a $\nabla_Q$-exact terms. 
This means that the energy-momentum tensor $T_{ij}$ associated to 
the target metric $G_{ij}$ is $\nabla_Q$-exact, 
then leading also to a topological field theory in the 
target space $M_T$ in the large limit $t\rightarrow\infty$.
Now we use \ll\ to argue that this topological character should hold for any
$t$.

\subsec{Observables.}

The observables ${\cal O} [\Phi]$ of any topological field theory are
metric-independent scalar objects belonging to the cohomology of $Q$:
\eqn\tara{
{\cal O}[\Phi]\in {Ker(\delta_Q)\over {Im(\delta_Q)}}.  } Note that we
could replace $\delta_Q$ by $\nabla_Q$ in the previous (and following) 
expressions ($\delta_Q =\nabla_Q$ when acting on target scalars). The
numerator of \tara\ just says that any observable has to be invariant under
the symmetry of the theory ($Q$). The denominator tells that, due to \ini
, two observables differing by a $Q$-exact 
quantity lead to the same correlators (and then, have to be identified 
as observables). So far, we have specified two of the fields of 
$\Phi$: $X^i$ and $\chi^i$. With them we can already construct 
observables satisfying \tara . In \refs{\wittenb} the analysis for 
the case where the internal manifold was bidimensional ($d=2$) was 
studied. It is not difficult to generalize the arguments there for the 
general case $d=m$. The result is the following.
 Given a $a$-dimensional homology cycle 
$\gamma_{a(i)}\in H_a (M_I)$ 
($0\leq a\leq m$ and $i:1,..,b_a=dim (H_a(M_I))$) and a p-form $A(X)\in
H^p (M_T ;R)$ we define $W^{\gamma_{a(i)}}_A$ by:
\eqn\obs{
W^{\gamma_{a(i)}}_{A}[X,\chi]=\int_{\gamma_{a(i)}} {{\cal O}^{a}_{A}}
(X,\chi)
}
where the object ${\cal O}^{a}_{A}(X,\chi)$ is given by:
\eqn\onp{
{\cal O}^{a}_A (X,\chi)=
{p\choose a}A(X)_{i_1,..,i_p}dX^{i_1}\wedge\ldots\wedge dX^{i_a}\chi^{i_{a+1}}
\dots\chi^{i_p}\qquad\qquad 0\leq a\leq m.
}
${\cal O}^{a-1}_{A}(X,\chi)$ and ${\cal O}^{a}_{A}(X,\chi)$ are 
easily seen to be related 
by the so-called ``topological descendent equations'' (use \re\ and \onp\ 
to prove this):
\eqn\tde{
d{\cal O}^{a-1}_{A}(X,\chi)=\delta_Q ({\cal O}^{a}_A (X,\chi)).
}
Similar arguments to those of \refs{\wittenb} adapted to the 
present case show that $W^{\gamma_{a(i)}}_A [X,\chi]$ defines an  
observable ($\delta_Q W^{\gamma_{a(i)}}_A [X,\chi]=0$) which depends on the 
homology class of $\gamma_{a(i)}$ and not on the particular cycle  
chosen. Therefore, 
to define the observables \obs\ we can take any basis $\gamma_{a(i)}$ of 
$H_a (M_I)$. An observable \obs\ can then be constructed for any choice 
of the pair $(A(X),\gamma_{a(i)})$ where $A(X)\in H^p (M_T;R)$ and 
$\gamma_{a(i)}\in H_a (M_I)$.

The correlators of these observables have the form:
\eqn\corr{
<\prod_{A,\gamma_{a(i)}} W^{\gamma_{a(i)}}_{A}>_t
}
The analysis in the limit $t\rightarrow\infty$ 
(see the action \tth ) localizes these correlators  
on the moduli space $\cal M$ ($s^i =0$ and $H^{i}_{*} =\rho_{i}^{*}
\nabla_Q s^{i}_{*}=0$). We 
can not continue this analysis without specifying a concrete model, 
however, let us just mention that the role of fermionic zero modes 
($\rho_{i}^{*}\nabla_Q s^{i}_{*}=0$) 
are going to be essential to establish the selection rules for 
obtaining non-zero outputs from \corr . These selection rules are 
dictated by index theorems depending on the manifolds $M_I$, $M_T$ and 
the differential operator $D$ defining the section $s^{i}_{*}$ 
(\refs{\wittenb, \wittenc}).

\newsec{Examples.}

In this section we will apply our formulas to derive the action and 
$\delta_Q$-transformations of two well known examples: Topological 
Quantum Mechanics (\refs{\labastida}) and Type A Topological Sigma 
Models (\refs{\wittenb, \yo}).

\subsec{Topological Quantum Mechanics.} 
  
In this case $M_I$ is taken to be $S^1$:
\eqn\laba{
X: S^1\rightarrow M_T.
}
($X$ can be thought as elements of $\pi_1 (M_T)$). 
The map $X$ can be locally described by functions $X^i (\tau)$
($i:1,..,n$), 
being $\tau$ a coordinate in $S^1$. The section is taken to be:
\eqn\labb{
s^i (X(\tau))= {d\over {d\tau}}X^i +V^i(X)
}
where $V^i(X)$ is some smooth vector field on $M_T$. 
In this situation we have:
\eqn\labc{
\eqalign{
\nabla_Q s^i (X)&=\delta_Q s^i (X)+\Gamma^{i}_{jk} (X)\chi^j s^k (X)\cr
&={d\over {d\tau}}\chi^i +\partial_j V^i (X)\chi^j +
\Gamma^{i}_{jk} (X)\chi^j ({d\over {d\tau}}X^k +V^k (X))\cr
&={\cal D}^i_j \chi^j.\cr}
}
where we have used the notation of \refs{\labastida}:
\eqn\labd{
{\cal D}^i_j ={\delta}^i_j {d\over {d\tau}} +{dX^k\over {d\tau}}
\Gamma^{i}_{kj} +D_j V^i.
}
We take the metric $A_{ij}[\Phi]=G_{ij}(X)$ in \tth\ (then, using 
\bull\ we get $\nabla_Q A_{ij}[\Phi]=0$). 
Therefore, the action \tth\ reads in this case:
\eqn\alab{
\eqalign{
S_t [\Phi]=&t\int e(\tau)d\tau G_{ij}(X)\Bigl\{ H^i H^j-
{1\over 4}({ d\over {d\tau}}X^i +V^i(X))
({d\over {d\tau}}X^j +V^j(X)) \cr
&-\rho^i {\cal D}^j_k \chi^k 
-{1\over 2}R^{j}_{klm} (X)\chi^k \chi^l \rho^i\rho^m\Bigr\} .\cr}
}
($e(\tau)$ is the ``einbein'' for $S^1$ making \alab\ invariant under 
reparametrizations on $\tau$). This is precisely the action obtained 
in a different way in \refs{\labastida}. 
The $\nabla_Q$-transformations can be obtained from \qt . 
We get (we write only the two last $\nabla_Q$-transformations of \qt , 
because the two first remain the same):
\eqn\pirri{
\eqalign{
&\bullet \nabla_Q \rho^i =H^i -{1\over 2}({d\over {d\tau}}X^i +V^i(X)) \cr
&\bullet\nabla_Q H^i ={1\over 2}{\cal D}^i_j 
\chi^j + {1\over 2}R^{i}_{jkl}\chi^j\chi^k\rho^l. \cr}
} 
Also, the analog of \im\ for the present case is:
\eqn\ein{
\delta_Q e(\tau)=0
}

The analysis of observables and the computation of the partition 
function was done in \refs{\labastida} and we refer there for details.

\subsec{Type A Topological Sigma Model.}

The topological sigma models are well known to be derived from the 
$N=2$ supersymmetric sigma models by performing the so-called 
``topological twisting'' (\refs{\wittenb , \yo}). The two basic $N=2$ 
multiplets are the chiral and the twisted-chiral multiplets. Twisting 
the first one gives the topological sigma model of type A. Twisting the 
second one we get the topological sigma model of type B (\refs{\wittenb,
 \yo}). There are several differences between type A and B topological
sigma models. Perhaps, the most relevant one is that while the type 
A models can be generalized even to real manifolds (as we will see 
shortly), the type B models depend strongly on the complex structure 
and has been formulated so far only for Kahler manifolds. Also, type 
A and B models formulated on Calabi-Yau spaces 
are known to be related by mirror symmetry (\refs{\wittenc}).

Let us derive the type A topological sigma models by applying our 
formulas.
First, the internal manifold $M_I$ is taken to be a two-dimensional, 
compact, oriented  Riemann surface $\Sigma$  
(endowed with the metric $g_{\alpha\beta} (\sigma)$):
\eqn\aa{
X:\Sigma\rightarrow M_T.
}
Also, we will consider now the case where the target manifold $M_T$ is 
an Hermitian manifold equipped with a complex structure $J^i_{\,\, j} (X)$ 
and the Hermitian metric  $G_{ij} (X)$ (this means that $J_{ij}(X)=G_{ik}(X) 
J^k_{\,\, j}(X)=-J_{ji}(X)$). 
Second, the section is taken to be:
\eqn\ab{
s^i_{\alpha} =-2(\partial_{\alpha} X^i +\epsilon_{\alpha}^{\beta} J^i_{\,\, j} 
\partial_{\beta} X^i )
}
(factors are chosen to make contact with the notation of \refs{\yo})
and the ``metric'' $A_{ij}^{**}$ in \norm\ is given by:
\eqn\ac{
A_{ij}^{\alpha\beta}(X,\sigma)={1\over 4} g^{\alpha\beta}(\sigma) G_{ij}(X)
}
From \ab\ and \ac\ one quickly derives:
\eqn\ad{
A_{ij}^{\alpha\beta}(X,\sigma)s^i_{\alpha}s^j_{\beta} =
2( g^{\alpha\beta}G_{ij}\partial_{\alpha} X^i\partial_{\beta} X^j
+\epsilon^{\alpha\beta} J_{ij}\partial_{\alpha} X^i 
\partial_{\beta} X^j).
}
where we have used that, for Hermitian manifolds, $g^{\alpha\beta} 
\epsilon_{\alpha}^{\,\,\tau}\epsilon_{\beta}^{\,\,\mu}G_{ij}J^i_{\,\, p}
J^j_{\,\, k}=g^{\tau\mu}G_{pk}$. Also one gets easily:
\eqn\ae{
A^{\alpha\beta}_{ij}\rho^i_{\alpha}\nabla_Q s^j_{\beta}=
-{1\over 2}(g^{\alpha\beta}G_{ij}\rho^i_{\alpha}D_{\beta}\chi^j +
\epsilon^{\alpha\beta}D_k J_{ij}\rho^i_{\alpha}\chi^k\partial_{\beta} X^j
+\epsilon^{\alpha\beta} J_{ij}\rho^i_{\alpha}D_{\beta}\chi^j).
}
where $D_{\alpha}V^i$ is the pull-back of the covariant derivative 
on the target:
\eqn\zeven{
D_{\alpha}V^i=\partial_{\alpha}V^i +\Gamma^i_{jk}V^j\partial_{\alpha}X^k.
}
Moreover:
\eqn\af{
{1\over 2}A^{\alpha\beta}_{ij}R^{j}_{klm}\chi^k\chi^l\rho^i_{\alpha}
\rho^m_{\beta}={1\over 8}g^{\alpha\beta}R_{klim}\chi^k\chi^l\rho^i_{\alpha}
\rho^m_{\beta}.
}
Again, $\nabla_Q A^{\alpha\beta}_{ij}=0$ (\bull ). Substituting all 
this information in our expression \tth\ we obtain:
\eqn\ag{
\eqalign{
S_t [\Phi]=&-{1\over 2}t\int d^2\sigma \sqrt{g}\Bigl\{
g^{\alpha\beta}G_{ij}\partial_{\alpha} X^i\partial_{\beta} X^j+
\epsilon^{\alpha\beta} J_{ij}\partial_{\alpha} X^i 
\partial_{\beta} X^j \cr
&\qquad\qquad\qquad -g^{\alpha\beta}G_{ij}\rho^i_{\alpha}D_{\beta}\chi^j -
\epsilon^{\alpha\beta}D_k J_{ij}\rho^i_{\alpha}\chi^k\partial_{\tau} X^j
-\epsilon^{\alpha\beta} J_{ij}\rho^i_{\alpha}D_{\beta}\chi^j \cr
&\qquad\qquad\qquad -{1\over 2}g^{\alpha\beta}G_{ij}H^i_{\alpha}H^j_{\beta}
+{1\over 4}R_{klim}\chi^k\chi^l\rho^i_{\alpha}
\rho^m_{\beta}\Bigr\}.\cr}
}
This expression is slightly different from that on \refs{\yo}, the reason 
being that we have not demanded here the self-duality conditions on 
the $\rho^i_{\alpha}$ and $H^i_{\alpha}$ fields ($\rho^i_{\alpha}=
\epsilon^{\,\,\beta}_{\alpha}J^i_{\,\, j}\rho^j_{\beta}$ and 
$H^i_{\alpha}=\epsilon^{\,\,\beta}_{\alpha}J^i_{\,\, j}H^j_{\beta}$). 
This means, in particular, that the topological 
action \ag\ can not be obtained from 
a twisting of a $N=2$ supersymmetric sigma model. However, if we desire 
to make full contact with the action obtained from a twisted $N=2$ 
supersymmetry we see straightforwardly that, imposing selfduality 
on the field $\rho^i_{\alpha}$, \ae\ can be written as:
\eqn\ah{
A^{\alpha\beta}_{ij}\rho^i_{\alpha}\nabla_Q s^j_{\beta}=
-(g^{\alpha\beta}G_{ij}\rho^i_{\alpha}D_{\beta}\chi^j +{1\over 2}
\epsilon^{\alpha\beta}D_k J_{ij}\rho^i_{\alpha}\chi^k\partial_{\tau} X^j).
}
then, on a Kahler manifold (where $D_i J^j_{\,\, k}=0$)
 we recover exactly the expression for the type A topological 
sigma model (\refs{\wittenb , \yo}):
\eqn\agb{
\eqalign{
S_t [\Phi]=&-{1\over 2}t\int d^2\sigma \sqrt{g}\Bigl\{
g^{\alpha\beta}G_{ij}\partial_{\alpha} X^i\partial_{\beta} X^j+
\epsilon^{\alpha\beta} J_{ij}\partial_{\alpha} X^i 
\partial_{\beta} X^j \cr
&\qquad\qquad\qquad -2g^{\alpha\beta}G_{ij}\rho^i_{\alpha}D_{\beta}\chi^j 
-{1\over 2}g^{\alpha\beta}G_{ij}H^i_{\alpha}H^j_{\beta}
+{1\over 4}R_{klim}\chi^k\chi^l\rho^i_{\alpha}
\rho^m_{\beta}\Bigr\} .\cr}
}
However, is not necessary to impose selfduality to show that our 
action \ag\ is by its own topological both on the target $M_T$ and on
$\Sigma$ (due to our general considerations in section 4).

Let us now use \nqt\ to derive the $\nabla_Q$-transformations for 
this case (again, the first two transformations in \nqt\ remain the 
same and we just write the two last ones):
\eqn\een{
\eqalign{
&\bullet \nabla_Q \rho^{i}_{\alpha}=H^i_{\alpha}+
\partial_{\alpha} X^i +\epsilon_{\alpha}^{\beta} J^i_{\,\, j} 
\partial_{\beta} X^j \cr
&\bullet \nabla_Q H^i_{\alpha}=-D_{\alpha}\chi^i -
\epsilon^{\beta}_{\,\, \alpha}J^{i}_{\,\, j}D_{\beta}\chi^j
+{1\over 2}R^{i}_{jkl}\chi^j\chi^k\rho^{l}_{\alpha}.\cr}
}
The $\nabla_Q$-transformation of $\rho^i_{\alpha}$ in \een\ is easily seen to 
violate selfduality on no-Kahler manifolds. Actually, if:
\eqn\twee{
\rho^i_{\alpha}=\epsilon_{\alpha}^{\,\,\beta}J^{i}_{\,\, j}
\rho^{j}_{\beta}
}
then:
\eqn\drie{  
\nabla_Q (\rho^i_{\alpha})=\epsilon_{\alpha}^{\,\,\beta}J^{i}_{\,\, j}
\nabla_Q (\rho^{j}_{\beta})+\epsilon_{\alpha}^{\,\,\beta}
\chi^k D_k (J^{i}_{\,\, j})
\rho^{j}_{\beta}\equiv \epsilon_{\alpha}^{\,\,\beta}J^{i}_{\,\, j}
\nabla_Q (\rho^{j}_{\beta}) +A^{i}_{\alpha}.
}
and self-duality is violated by the $\nabla_Q$-transformations due 
to the term $A^{i}_{\alpha}\equiv\epsilon_{\alpha}^{\,\,\beta}
\chi^k D_k (J^{i}_{\,\, j})\rho^{j}_{\beta}$. This term is obviously 
zero for a Kahler manifold ($D_k J^i_j =0$) but not for an arbitrary 
Hermitian manifold. However, we can check that $A^i_{\alpha}$ is 
anti-selfdual:
\eqn\vier{
A^i_{\alpha}=-\epsilon_{\alpha}^{\,\,\beta}J^{i}_{\,\, j}
A^{j}_{\beta}
}
and consequently, from \drie , we could naturally define a new operator:
\eqn\vijf{
\hat{\nabla}_Q \rho^i_{\alpha}=\nabla_Q \rho^i_{\alpha}-
{1\over 2}A^{i}_{\alpha}
}
which is selfdual. From this we could define,
if we want, a selfdual $\hat{Q}$-transformation 
(\refs{\yo}) for Hermitean  (non-Kahler) manifolds (note that the action 
\ag\ changes also of form if we define it $\hat{\nabla}_Q$-exact). But 
we stress again that self-duality is not a necessary condition to have 
a topological theory in our formalism.

Let us now argue that the theories defined through the actions \ag\ and \agb\ 
do not depend on the complex structure $J^i_{\,\, j}$. The argument 
is completely similar to the one we used on section $4$ to show that 
the theory is topological on the target manifold $M_T$. 
First we note that the $\nabla_Q$-transformations in \een\ depend
on the complex structure $J^i_{\,\, j}$ through the section $s^i_{\alpha}
 (X,J(X))$. Due to the 
$\nabla_Q$-exactness of the action we can analyze the theory on the 
large $t$ limit. There we have localization on the instanton configurations
$s^i_{\alpha}=0$ and then all the dependence of \een\ in $J^i_{\,\, j}$ 
disappears. In this case, deformations of the complex structure and 
$\nabla_Q$ transformations commute and the theory is invariant under 
that deformations of $J^i_{\,\, j}$. The introduced reader 
could be surprised at this point, because the type A topological theories  
formulated on Kahler manifolds ($dJ=0$ where $J=J_{ij}dX^i\wedge dX^j$ 
and $J_{ij}=G_{ik}J^{k}_{\,\,j}$) 
are known to depend on the Kahler class $J$. 
But we remark here that the action employed by Witten $S^W_t [\Phi]$
in \refs{\wittenc} is 
not \agb\ but:
\eqn\pablob{
\eqalign{
S^W_t [\Phi]=&S_t [\Phi]-t\int{d^2 \sigma \sqrt{g(\sigma)} X^{*}(J)}\cr
=&t \nabla_Q (\int{d^2 \sigma\sqrt{g(\sigma)}(\rho^{*}_{i} [\Phi] s^{i}_{*}
[\Phi]}+\rho_{i}^{*}\nabla_Q \rho^{i}_{*})) 
-\int{d^2 \sigma \sqrt{g(\sigma)} X^{*}(J)}\cr \equiv &S_t[\Phi] -K(J).\cr}
}
being:
\eqn\pull{
K(J)\equiv\int{d^2 \sigma \sqrt{g(\sigma)} X^{*}(J)}=\int{
d^2 \sigma \sqrt{g(\sigma)} 
\epsilon^{\alpha\beta}J_{ij}\partial_{\alpha} X^i\partial_{\beta} X^j}
}
($X^{*}(J)$ is the pullback of the Kahler form). 
This term is seen to be invariant 
under $\nabla_Q$-transformations:
\eqn\gil{
\nabla_Q \Bigl(\epsilon^{\alpha\beta} J_{ij}\partial_{\alpha}X^i 
\partial_{\beta} X^j\Bigr)=2\epsilon^{\alpha\beta}D_{\alpha}(J_{ij}
\chi^i\partial_{\beta} X^j)
}
(we have used the Kahler condition 
$D_i J^k_{\,\, l}=0$), but is not $\nabla_Q$-exact. 
Therefore the topological character of the theory  
is not guaranteed (we can not use the large $t$ 
limit if the action is not $\delta_Q$ or $\nabla_Q$-exact). In fact,  
the term \pull\ added in the action \pablob\ is a topological invariant 
(and consequently, the theory defined by the action \pablob\ is 
still invariant under deformations of the target metric $G_{ij} (X)$
and the internal metric $g_{\mu\nu}$), 
however, \pull\ depends on the homotopy class of the map $X$ and the 
cohomology class of the closed form $J$ (and then, is sensible to changes
of the Kahler form $J(X)=J_{ij}dX^i\wedge dX^j$). As a consecuence, 
the theory defined by $S^W_t$ depends on the Kahler form $J$ whereas the
one defined by $S_t$ does not.

Before finishing this section, let us do some general comments. The type A 
topological action that we have constructed here is exactly the same 
as the one that is obtained by twisting the $N=2$ supersymmetric 
sigma models (when the manifold $M_T$ is Kahler
and we demand self-duality conditions on the fields $\rho^i_{\alpha}$
and $H^i_{\alpha}$). The generalization 
to Hermitian manifolds was done by Witten in \refs{\wittenb}. In the 
present context we see that the type A topological sigma models can 
be generalized even to real manifolds. One just has to take the section 
\ab\ without the complex structure term and use our formulas \tth\ and 
\qt\ to derive the acction and $Q$-transformations. 
With respect to the observables of 
the theory, they  were analyzed in \refs{\wittenc}. 
Here we note from \ab\ that 
the localization of correlators are going to take place on the moduli 
space of holomorphic instantons:
\eqn\zes{
-{1\over 2}s^i_{\alpha}= \partial_{\alpha}X^i+\epsilon_{\alpha}^{\,\,\beta}
J^i_{\,\, j}\partial_{\beta} X^j =0.
}
One has to study again carefully the fermionic zero modes to know 
which of the observables \onp\ give non-trivial correlators. 
This depends on $\Sigma $, $M_T$ and index theorems. We refer 
to \refs{\wittenc} for details. The result is that, on Kahler manifolds, 
 correlators split on an addition $\sum_k$ of 
 intersection forms on the moduli space of holomorphic instantons
of degree $k$ (over instantons of the type \zes\ we have that $K(J)=k$ where,
using a proper normalization, $k$ is an integer 
\refs{\wittenc}).  
If we use the action $S^W_t$ (\pablob ) instead of $S_t$ (\agb ) these 
intersection numbers are weighted by exponentials of the degree 
of the corresponding holomorphic instantons ($e^{-itK(J)}\sim e^{-itk}$). 
In the case that we work 
with a real manifold $M_T$ the generalization is straightforward with 
our formalism. We can just the section $s^i_{\alpha}$ to
be \ab\ without the complex structure term and substitute in our formulae
\tth\ and \qt\ to get the action and transformations. A simple analysis 
gives that correlators, in this case,  are 
just the classical intersection numbers of submanifolds of $M_T$ 
which are Poincar\'e duals to the corresponding forms $A(X)$ entering in 
the observables \onp\ in the correlators (note that the solutions 
of the moduli equations $\partial_{\alpha} X^i =0$ are the constant 
maps and therefore, $\cal M$ coincides with the target manifold $M_T$).

\newsec{Conclusions}

We have introduced a wide class of topological field theories of 
maps $X^i :M_I\rightarrow M_T$ from a $m$-dimensional internal manifold
$M_I$ to a $n$-dimensional target manifold $M_T$ localizing the 
correlators on a desired moduli space of instantons $\cal M$. The minimal 
action is given in \tth\ and the general $Q$-transformations are given in  
\nqt . To guarantee the topological character of the theory we 
were forced to demand the $\rho^{i}_{*} [\Phi]$ fields to be independent
both of the internal metric $g_{\mu\nu}$ and the target metric $G_{ij}$.
Our models contain previously known topological systems as 
particular cases, like 
the topological sigma models of type A. However our approach does not 
involve the twist procedure of an $N=2$ supersymmetric sigma model. 
This allowed us to formulate topological matter of type A in 
real manifolds (to our knowledge, so far, they have been formulated 
only for Hermitean manifolds \refs{\wittenb}). 
In the case of real manifolds, when the section is chosen to be 
$s^i_{\alpha}=\partial_{\alpha} X^i$, the correlators 
turn out to be classical intersection numbers of submanifolds 
of the target space $M_T$. 

It would be interesting to study if the models here introduced 
contains the other type of known topological sigma models (type B). 
Also our formalism could be generalized to the case in which the 
topological charge $Q$ is not nilpotent but 
closses on a group of symmetry transformations
of the theory.

\bigskip
{\bf Acknowledgements}

We would like to thank J.M.F. Labastida and A.V. Ramallo for valuable 
discussions. This work is mainly supported by a Spanish Government 
grant and partially supported by DOE-91ER40618 and DGICYT (PB93-0344).

\listrefs
\end